\documentclass[prl,preprint,showpacs,amsmath,amssymb,superscriptaddress,letterpaper]{revtex4}

\usepackage{graphicx}
\usepackage{bm}

\begin{document}

\title{Point singularities and suprathreshold stochastic resonance in optimal coding}

\author{Mark D. McDonnell}
 \email{mmcdonne@eleceng.adelaide.edu.au}
\affiliation{ School of Electrical and Electronic Engineering \&
Centre for Biomedical Engineering, The University of Adelaide, SA
5005, Australia }
\author{Nigel G. Stocks}
 \email{ngs@eng.warwick.ac.uk}
\affiliation{School of Engineering, The University of Warwick,
Coventry CV4 7AL, United Kingdom }
\author{Charles E.M. Pearce}
 \email{cpearce@maths.adelaide.edu.au}
\affiliation{ School of Mathematical Sciences, The University of
Adelaide, SA 5005, Australia }
\author{Derek Abbott}
 \email{dabbott@eleceng.adelaide.edu.au}
\affiliation{ School of Electrical and Electronic Engineering \&
Centre for Biomedical Engineering, The University of Adelaide, SA
5005, Australia }

\date{\today}

\begin{abstract}
Motivated by recent studies of population coding in theoretical
neuroscience, we examine the optimality of a recently described
form of stochastic resonance known as suprathreshold stochastic
resonance, which occurs in populations of noisy threshold devices
such as models of sensory neurons. Using the mutual information
measure, it is shown numerically that for a random input signal,
the optimal threshold distribution contains singularities. For
large enough noise, this distribution consists of a single point
and hence the optimal encoding is realized by the suprathreshold
stochastic resonance effect. Furthermore, it is shown that a
bifurcational pattern appears in the optimal threshold settings as
the noise intensity increases. Fisher information is used to
examine the behavior of the optimal threshold distribution as the
population size approaches infinity.
\end{abstract}


\maketitle

A fascinating aspect of the behavior of populations of neurons is
their capability of reliability in the presence of very low signal
to noise ratios~\cite{DeWeese.95}.  It has been established by
many studies that improved performance in individual neurons can
be achieved in the presence of large ambient noise, by a mechanism
known as Stochastic Resonance
(SR)~\cite{Wiesenfeld.94,Levin.96,Gammaitoni.Jan98}. Motivated by
the established fact that a single stimulus can induce a response
in many sensory neurons, we aim to link SR with a major branch of
theoretical neuroscience, that of population coding, in which one
of the aims is to understand how populations of neurons encode
input stimuli~\cite{Pouget.03}. The approach we take is to analyze
the optimal encoding of a random input signal by a population of
simple threshold devices, or comparators, all of which are subject
to additive {\it iid} input noise. Although this approach greatly
simplifies the dynamics of realistic neural models, it does
encapsulate the main nonlinearity: that of a threshold that
generates an output spike when crossed. A natural measure to use,
and one which recently has been used extensively in computational
neuroscience~\cite{DeWeese.95}, is that of mutual information.

The vast majority of studies on SR in static nonlinearities and
neurons have been restricted to the case of subthreshold signals
since, for a single device and suprathreshold stimuli, noise
enhanced signal transmission disappears. Hence, it has been
pointed out that stochastic resonance is a sub-optimal means for
improving system performance, since optimal performance can be
gained by adjusting the threshold
value~\cite{DeWeese.95,Abbott.01}. However, we show here that this
does not necessarily apply to systems consisting of more than one
threshold device receiving the same input signal and subject to
independent noise. Previously, it has been shown for
suprathreshold signal levels in such a system that the mutual
information between the input and output signals has a maximum
value for nonzero noise intensity. This phenomenon was termed
Suprathreshold Stochastic Resonance (SSR) to illustrate the fact
that it is a form of stochastic resonance that is not restricted
to subthreshold signals~\cite{Stocks.Mar2000}. Subsequently, the
effect was also shown to occur in FitzHugh-Nagumo model
neurons~\cite{Stocks.01c} and applied to cochlear implant
encoding~\cite{Stocks.02}.

Here for the first time, we discuss the optimality of SSR by
examining whether the mutual information can be increased by
adjusting the thresholds, which we denote as $\{\theta_n\},
n=1,..,N$, while keeping the noise constant. Using a Gaussian
signal and \textit{iid} Gaussian noise, we show numerically that
above a certain noise intensity the optimal threshold settings
occur when all thresholds are equal to the signal mean. Hence, we
show the SSR effect is a case of SR where making use of the
ambient noise truly provides the optimal system response.
Furthermore, we show that as noise increases from zero, (where the
optimal thresholds are widely distributed across the dynamic range
of the signal), the values of the optimal threshold settings go
through a number of transitions where more and more threshold
values accumulate to fewer and fewer points, in a series of
bifurcations. Such an accumulation of optimal thresholds to a
small number of point singularities appears to persist if we let
$N$ approach infinity and hence, in this case there must be a
transition from continuous to singular solutions of the optimal
threshold values.

\begin{figure}
\begin{center}\includegraphics[width=1.0\textwidth]{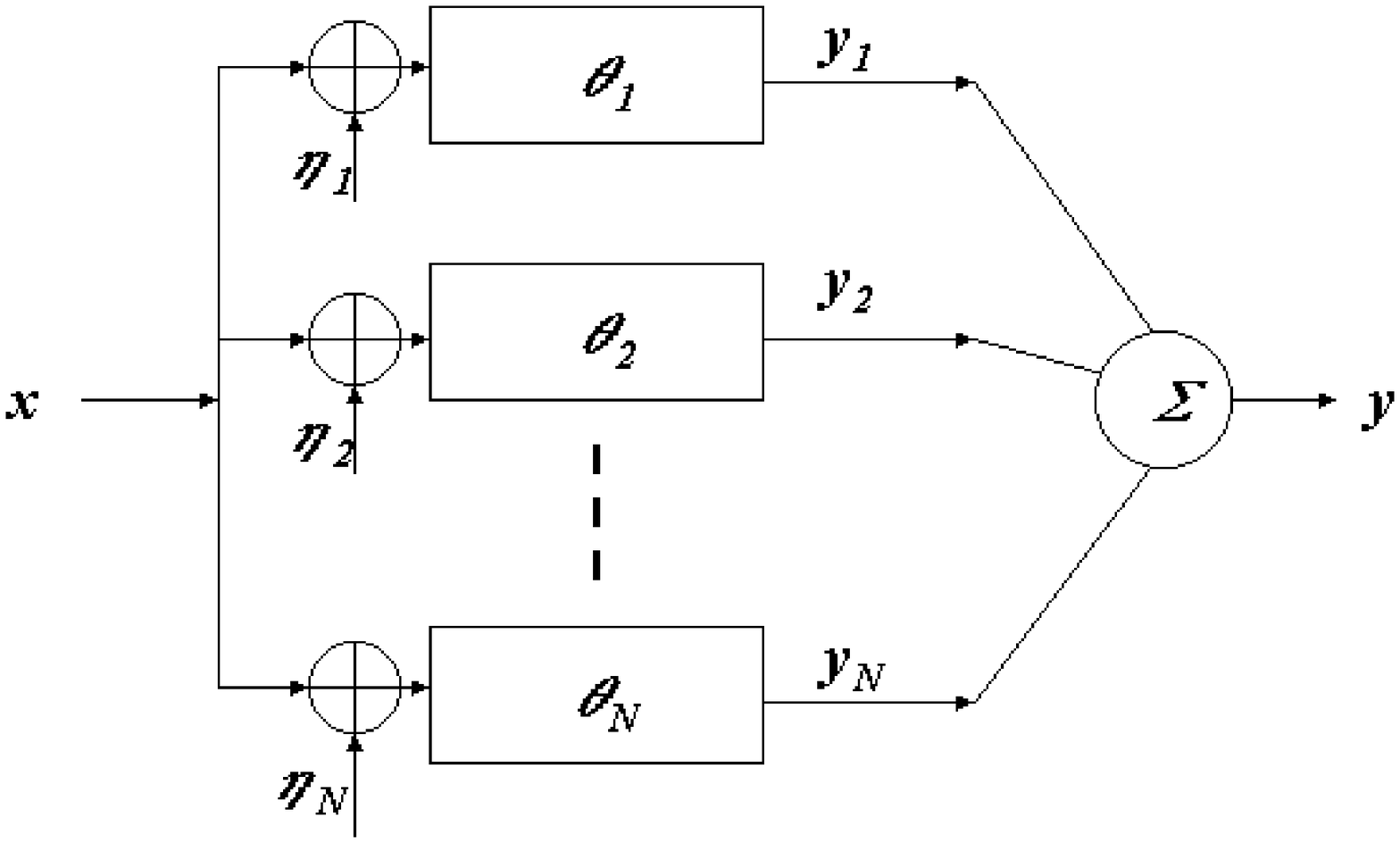}\end{center}
\caption{\label{f:array}Array of $N$ noisy threshold devices. Each
device receives the same input signal sample, and is subject to
independent additive noise. The output from each device is unity
if the sum of the signal and noise at its input is greater than
the corresponding threshold and zero otherwise. The overall
output, $y$, is the sum of the individual outputs.}
\end{figure}

The model in which SSR occurs is shown in Fig.~\ref{f:array}. It
consists of $N$ threshold devices all receiving the same sample of
a random input signal $x$, with pdf $P(x)$ where the $n$--th
device is subject to continuously valued \textit{iid} additive
noise, $\eta_{n}$ ($ n = 1,..,N$) with pdf $R(\eta)$, which is
also independent of $x$. The output from each comparator, $y_n$,
is unity if the input signal plus the noise is greater than the
threshold, $\theta_{n}$, of that device and zero otherwise. The
outputs from each comparator are summed to give the overall output
signal, $y$. Hence, $y$ is a discrete signal taking on integer
values from $0$ to $N$ and is a nondeterministic and lossy
encoding of the input signal. If we impose the constraint that all
thresholds are set to the same value, the SSR effect can occur, in
which case the mutual information has a maximum for nonzero noise
intensity~\cite{Stocks.Mar2000}. The effect is maximized when the
thresholds are all set to the signal mean~\cite{Stocks.01b}. Such
behavior has also been shown with other
measures~\cite{McDonnell.02MEJ,McDonnell.02FNL,Rousseau.03b}.

Our objective here however, is to relax the constraint that all
thresholds are identical, and to find the threshold values,
$\{\theta_n^*\}$, that maximize the mutual information. This is
expressed as $I(x,y) = -\sum_{n=0}^N
P_y(n)\log_{2}P_y(n)-\left(-\int_{-\infty}^\infty
P(x)\sum_{n=0}^NP(y=n|x)\log_{2}P(y=n|x)dx\right)$, where $P(x)$
is assumed known and we have $P_y(n) = \int_{-\infty}^\infty
P(y=n|x)P(x)dx$. Given this, the mutual information depends only
on the conditional probability of the output given the input,
$P(y|x)$. Let $\hat{P}_n(x)$ be the probability of device $n$
being ``on'' (that is, signal plus noise exceeding the threshold
$\theta_{n}$), for a given value of input signal $x$. Then
\begin{equation}\label{Prob_on}
    \hat{P}_n(x) = \int_{\theta_{n}-x}^\infty R(\eta)d\eta = 1-F_R(\theta_n-x),
\end{equation}
where $F_R$ is the cumulative distribution function (cdf) of the
noise and $n =1,..,N$. For the particular case when the thresholds
all have the same value, then each $\hat{P}_n(x)$ has the same
value for all $n$ and we have $P(y|x)$ given by the binomial
distribution~\cite{Stocks.Mar2000}. However, in general it is
difficult to find analytical expressions for $P(y|x)$ and we will
rely on numerics. Given any arbitrary $N$, $R(\eta)$, and
$\{\theta_n\}$, $\{\hat{P}_n(x)\}$ can be calculated exactly for
any value of $x$ from (\ref{Prob_on}), from which $P(y|x)$ can be
found using an efficient recursive formula~\cite{McDonnell.02MEJ},
and hence the mutual information calculated numerically. Our
problem of finding the threshold settings that maximize the mutual
information can now be expressed as a nonlinear optimization
problem, where the cost function to maximize is the mutual
information, and there are structural constraints on how $P(y|x)$
is obtained,
\begin{eqnarray}\label{Opt_theta}
    & \mbox{Find:}\quad& \max_{P(y|x)}I(x,y)\nonumber\\
&\mbox{subject to:}& P(y|x)\quad\mbox{is a function of}\quad\{\hat{P}_n(x)\},\nonumber\\
 &&\sum_{n=0}^NP(y=n|x)=1\quad\forall x,\nonumber\\
 &&\mbox{and}\quad\{\theta_n\} \in \mathbf{R^n}.
\end{eqnarray}
This formulation is similar to previous work on clustering and
neural coding problems solved using a method known as
deterministic annealing~\cite{Rose.90,Rose.98,Dimitrov.01}. In
particular, the formulation reached in~\cite{Dimitrov.01} can be
expressed in a fashion identical to (\ref{Opt_theta}) with the
structural constraints removed. Hence, the solution method used in
that work to find the optimal conditional distribution, $P(y|x)$,
cannot be used here, and instead we concentrate on optimizing the
only free variable, the set $\{\theta_n\}$.  To this end, we have
successfully applied a random search method based on simulated
annealing to find near optimal solutions to (\ref{Opt_theta}) for
any given $N$, $P(x)$ and noise intensity. We have found this
method to be highly efficient and effective and have tested it by
using other solution methods that do not rely on any assumptions
of local convexity in the solution space, including a genetic
algorithm. These methods always provide very nearly the same
solution as our method, but are far slower to converge.

\begin{figure}
\begin{center}\includegraphics[width=1.0\textwidth]{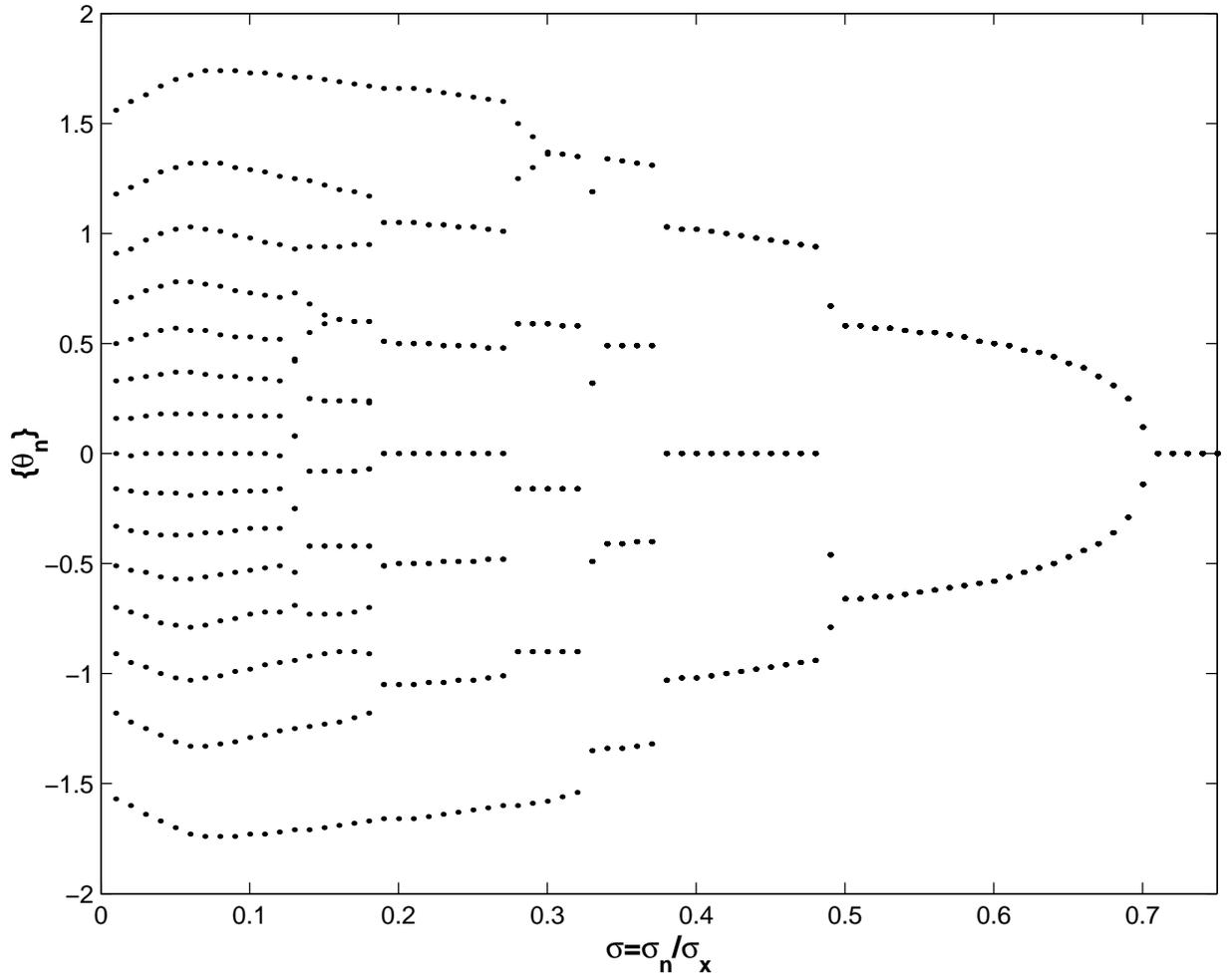}\end{center}
\caption{\label{f:N15}Plot of optimal thresholds, $\{\theta_n^*\}$
against $\sigma$ for $N=15$ and Gaussian signal and noise.}
\end{figure}

We present results for the case of \textit{iid} zero mean Gaussian
signal and noise distributions. If the noise has variance $\sigma
_\eta^2$ then we have $\hat{P}_n(x) =
\frac{1}{2}+\frac{1}{2}\mbox{erf}\left(\frac{x-\theta_n}{\sqrt{2}\sigma_\eta}\right)$,
where $\mbox{erf}$ is the error function. Let $\sigma =
\sigma_\eta/\sigma_x$ where $\sigma_x^2$ is the variance of the
Gaussian signal. Note that for the case of $\sigma_\eta=0$ that it
possible to analytically determine the optimal
thresholds~\cite{McDonnell.02MEJ}, and that in this case, each
threshold has a unique value. Fig.~\ref{f:N15} shows our results
for the optimal threshold settings for $N=15$ plotted against
increasing $\sigma$. Several interesting features are present in
these results.

\begin{figure}
\begin{center}\includegraphics[width=1.0\textwidth]{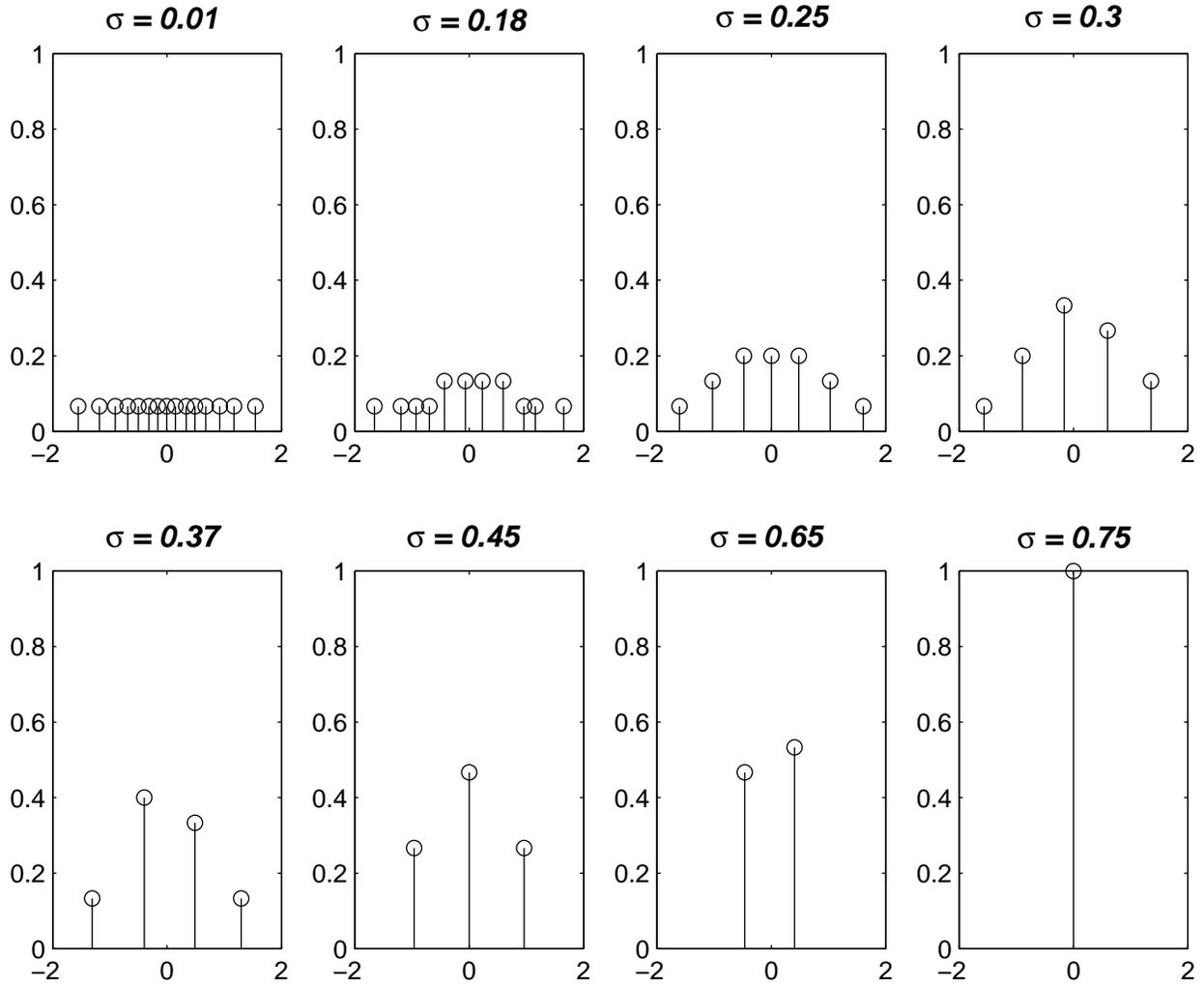}\end{center}
\caption{\label{f:thetaDist}Threshold point density function,
$\lambda(x,\sigma)$ obtained for $N=15$ and various values of
$\sigma$. The y-axes give the fraction of thresholds at each
accumulation point, $v(j,\sigma)$ and the x-axis gives the
threshold values, $x=\Theta_j$.}
\end{figure}

Firstly, for increasing noise, the optimal thresholds cluster to
particular values, which we will denote as an accumulation point.
There are also bifurcations for increasing noise. At each
bifurcation, the number of accumulation points decreases, so that
overall, the noise intensity is divided into $m \le N$ distinct
regions, each with $k_m$ accumulation points. The fraction of
thresholds at each accumulation point is shown in
Fig.~\ref{f:thetaDist} for various values of $\sigma$. For noise
greater than the final bifurcation point, we have the SSR region
occurring, that is, the optimal solution is for all thresholds to
be equal to the signal mean of zero. It can be seen that sometimes
a continuous bifurcation occurs and one accumulation point splits
into two (as noise decreases), and on other occasions a
discontinuous bifurcation occurs.

We also note that there are regions where asymmetry occurs about
the $x$-axis. Furthermore, our results have also shown the
existence of at least two identical global optimal threshold
settings for certain values of $\sigma$. This can occur at a
bifurcation point, where two different sizes of accumulation
points can provide the same mutual information, or between
bifurcations where the two solutions are sets of thresholds that
are simply the negative of each other. The bifurcational structure
is quite surprising, but appears to be fundamental to the problem
type, as we have found similar bifurcation structures in the
optimal threshold setting for measures other than mutual
information, including correlation coefficient and mean square
distortion (error variance), and other signal and noise densities.
Finally, it is evident that above a certain value of $\sigma$ the
SSR situation is optimal. That is, the optimal quantization for
large noise is to set all thresholds to the signal mean. This
result shows for the first time that stochastic resonance can be
optimal in threshold systems.

To describe our results mathematically, define $N$ states with
labels $z=n/N$. Let the set of $N$ optimal thresholds, be ordered
by increasing value to get the sequence $(\theta_z^*)_{z=0}^1$. In
the absence of noise, it is straightforward to show that each
optimal threshold is given by $\theta_z^*= F_x^{-1}\left(z\right)$
where $F_x^{-1}(.)$ is the inverse cdf of the signal distribution.

We introduce a concept used in the theoretical analysis of high
resolution quantizers in information theory: that of a quantizer
point density function, $\lambda(x)$, defined over the same
variable as the source~\cite{Gray.98}. The point density function
has the property that $\int_x\lambda(x)dx=1$, and gives the
density of thresholds across the signal dynamic range. For nonzero
noise, we observe that for a given value of $\sigma$, our
empirically optimal $(\theta_z^*)_{z=0}^1$ can take on at most
$k(\sigma)$ unique values. As $\sigma$ increases, $k$ decreases at
each bifurcation. Denote $v(j,\sigma)$ as the fraction of
thresholds assigned to the $j$-th accumulation point at noise
intensity $\sigma$ where $j\in\{1,..,k(\sigma)\}$, so that
$\sum_{j=1}^{k(\sigma)}v(j,\sigma)$=1. Denote the value of
$\theta_z^*$ for accumulation point $j$ as $\Theta_j$, so that
there are $Nv(j,\sigma)$ thresholds at $x=\Theta_j$. Hence, we can
write a point density function as a function of $\sigma$ as
\begin{equation*}\label{density}
    \lambda(x,\sigma)=\sum_{j=1}^{k(\sigma)}v(j,\sigma)\delta(x-\Theta_j).
\end{equation*}
We also note that $\int_{x=-\infty}^{a}\lambda(x,\sigma)dx$ is the
fraction of thresholds with values less than or equal to $a$. For
the special case of $\sigma=0$ we can write the analytically
optimal point density function as
\begin{eqnarray*}\label{infdensity}
  \lambda(x,0) =  \sum_{j=1}^{N}v(j,0)\delta(x-\Theta_j)
  = \sum_{n=1}^N\frac{1}{N}\delta\left(x-F_{x}^{-1}\left(\frac{n}{N}\right)\right).
\end{eqnarray*}

We can make some analytical progress on the solution to
(\ref{Opt_theta}) by allowing the population size $N$ to approach
infinity, a case which is biologically relevant~\cite{Wu.02}. Let
$N\rightarrow\infty$, so that $z$ is continuous in the region
$z\in[0,1]$. We can then say that the sequence of optimal
thresholds $(\theta_z^*)_{z=0}^1$ defines a strictly
non-decreasing, function $\Theta(z)$ defined on the continuous
interval $z\in[0,1]$. For the noiseless case, we have
$\Theta(z)=F_x^{-1}\left(z\right)$, which is a continuously valued
function on $z\in[0,1]$. For any continuously valued pdf, $P(x)$,
there is a one-to-one mapping from $z$ to the support of $P(x)$.
Furthermore,
\begin{equation*}\label{infdensity1}
  \lim_{N\to\infty}\lambda(x,0)
  = \int_{z=0}^1\delta\left(x-F_{x}^{-1}(z)\right)dz=P(x),
\end{equation*}
that is, the point density function is the pdf of
the signal.

However, for nonzero noise, our numerical solutions of
(\ref{Opt_theta}) indicate that even for very large $N$ the
accumulation points and bifurcational structure persists. Hence,
if we assume that this is the case also for infinite $N$, there
must be a transition at some $\sigma$ from a continuously valued
to a discretely valued optimal $\Theta(z)$. This is the reason
that we claim the optimal threshold distribution contains point
singularities for $\sigma>0$.

Furthermore, our numerical results indicate that the location of
the $m$-th bifurcation tends to converge to the same value of
noise as $N$ increases. Under the assumption that this holds for
infinite $N$, we are able to make use of an approximation to the
mutual information to find the location of the final bifurcation,
that is the smallest noise intensity for which SSR is the optimal
coding strategy. This approximation relies on an expression for a
lower bound on the mutual information involving the Fisher
information, $J(x)$, and the entropy of an efficient estimator for
$x$. Fisher information has previously been studied in the context
of SSR~\cite{Rousseau.03b} and is a measure of how well the input
signal, $x$, can be estimated from a set of $N$ observations. In
the limit of large $N$, the entropy of an efficient estimator
approaches the entropy of the input signal, and if the
distribution of this estimator is Gaussian then the lower bound
becomes asymptotically equal to the actual mutual information
as~\cite{Brunel.98}
\begin{equation}\label{MI}
    I(x,y) =
    H(x)-0.5\int_xP(x)\log_2{\frac{2{\pi}e}{J(x)}}dx.
\end{equation}
The Fisher information for the system in Fig.~\ref{f:array} is the
same regardless of whether the $N$ devices are summed or not.
Hence the Fisher information can be expressed as the sum of the
$N$ individual Fisher informations~\cite{Cover} as
\begin{equation*}\label{Fisher_arb2}
    J(x) =
    \sum_{n=1}^N\frac{\left(\frac{d\hat{P}_n(x)}{dx}\right)^2}{\hat{P}_n(x)(1-\hat{P}_n(x))},
\end{equation*}
which for the SSR case is identical to the
expression derived for the Fisher information in~\cite{Hoch.03a}.
For a zero mean Gaussian input signal (\ref{MI}) becomes
\begin{equation}\label{MI1}
    I(x,y) =
    0.5\int_xP(x)\log_2{\left(\frac{\sigma_x^2}{J(x)}\right)}dx,
\end{equation}
and it is now possible to efficiently solve (\ref{Opt_theta}) for
large $N$. However we note that the condition of the distribution
of $y$ given $x$ to be approximately Gaussian is only true for
noise in the vicinity, or larger, than the final bifurcation. In
this region, our empirical results for (\ref{Opt_theta}) show that
$v(1,\sigma)=v(2,\sigma)=0.5$, and that
$\lambda(x,\sigma)=0.5\delta(x-t)+0.5\delta(x+t)$, where
$t{\ge}0$. Under the assumption that this holds for very large
$N$, it is straightforward to numerically find the value of $t$
that maximizes (\ref{MI1}) for any given $\sigma$. This
maximization finds that the asymptotic location of the first
bifurcation is at $\sigma \simeq 0.91$.

To summarize, we have shown that the optimal encoding of a
Gaussian input signal by an array of noisy threshold devices
contains point singularities in its threshold distribution, the
number of which decreases in a series of bifurcations as the noise
intensity increases. We have also found that for large enough
noise, the optimal encoding is for all thresholds to be equal to
the signal mean. This shows that SSR is a form of threshold-based
SR that can be optimal. Finally, a Fisher information approach has
shown that for very large population sizes, and Gaussian signal
and noise, the noise intensity at which SSR becomes optimal
converges to $\sigma \simeq 0.91$.

\begin{acknowledgments}
This work was funded by the Australian Research Council, The
Australian Academy of Science, the Cooperative Research Centre for
Sensor Signal and Information Processing (Australia), and the
Leverhulme Trust, grant number F/00 215/J and we thank them for
their support.
\end{acknowledgments}

\end{document}